# A Classical Fuzzy Approach for Software Effort Estimation on Machine Learning Technique

S.Malathi[1] and Dr.S.Sridhar[2]

[1] Research Scholar, Department of CSE, Sathyabama University
Chennai, Tamilnadu, India

[2] Research Supervisor, Department of CSE and IT, Sathyabama University
Chennai, Tamilnadu, India

**Abstract**
Software Cost Estimation with resounding reliability, productivity and development effort is a challenging and onerous task. This has incited the software community to give much needed thrust and delve into extensive research in software effort estimation for evolving sophisticated methods. Estimation by analogy is one of the expedient techniques in software effort estimation field. However, the methodology utilized for the estimation of software effort by analogy is not able to handle the categorical data in an explicit and precise manner. A new approach has been developed in this paper to estimate software effort for projects represented by categorical or numerical data using reasoning by analogy and fuzzy approach. The existing historical datasets, analyzed with fuzzy logic, produce accurate results in comparison to the dataset analyzed with the earlier methodologies.

## 1. Introduction

The software environment has evolved significantly in the last 30 years. To estimate software development effort, the use of the neural networks has been viewed with skepticism by majority of the cost estimation community. Even though neural networks have exposed their strengths in solving multifarious problems, their limitation of being 'black boxes' has limited their usage as a common practice for cost estimation [4]. Some models carry a few advantageous features of the neuro-fuzzy approach, such as learning capability and excellent interpretability, while maintaining the qualities of the COCOMO model [6].

Estimation by Analogy is simple and flexible, compared to algorithmic models. Analogy technique is applied effectively even for local data which is not supported by algorithmic models [2], [8]. It can be used for both qualitative and quantitative data, reflecting closer types of datasets found in real life. Analogy based estimation has the potential to mitigate the effect of outliers in a historical data set, since estimation by analogy does not rely on calibrating a single model to suit all the projects. Unfortunately, it is difficult to assess the preliminary estimation as the available information about the historic project data during early stages is not sufficient [11]. The proposed method effectively estimates the software effort using analogy technique with the classical fuzzy approach.

The research paper is organized as follows: Section 2 deals with some of the recent research works related to the proposed technique. Section 3 describes the proposed technique and Section 4 discusses about the experimentation and comparative results with necessary tables and graphs and Section 5 concludes the paper.

## 2. Related Work

The proposed technique elucidates the effective estimation of effort. Several researchers have carried out researches in the field of effort estimation for the software projects using various techniques [9]. A few of the significant researches have been highlighted here for iris recognition.

Fuzzy logic has been applied to the COCOMO using membership functions such as Symmetrical Triangles and Trapezoidal Membership Function (TMF) to signify the cost drivers. The limitation of the latter function is that a few attributes were assigned the maximum degree of compatibility instead of lower degree. To overcome this drawback, Ch. Satyananda Reddy et al. [7] proposed the usage of Gaussian Membership Function (GMF) for the cost drivers by studying the behaviour of COCOMO cost drivers. COCOMO dataset has been used in the proposed methodology and the experiments envisage the scientific approach and compare the same with the standard version of the COCOMO. It is adduced that the Gaussian function performs better than the trapezoidal function as the latter facilitates a smooth transition in its intervals and the achieved results are closer to the actual effort.

Ahmeda and Muzaffar [6] dealt with the imprecision and uncertainty in the inputs of effort prediction. M.Kazemifard et al. [1] uses a multi agent system for handling the characteristics of the team members in fuzzy system. There are many studies that utilized the fuzzy systems to deal with the ambiguous and linguistic inputs of software cost estimation [3].

Wei Lin Du et al. [5] proposed an approach combining the neuro-fuzzy technique and the SEER-SEM effort estimation algorithm. The continuous rating values and linguistic values are the inputs of the proposed model for avoiding the deviation in estimation among similar projects. The performance of the proposed model has been improved by designing and evaluated with data from published historical projects. The evaluation results indicate that the estimation with the proposed fuzzy model containing analogy reasoning produce better results in comparison with the existing estimated results [4] that uses feature selection algorithm.

## 3. Proposed Methodology

### 3.1 Effort Estimation

Fuzzy logic is based on human behaviour and reasoning. It has an affinity with fuzzy set theory and applied in situations where decision making is difficult. A Fuzzy set can be defined as an extension of classical set theory by assigning a value for an individual in the universe between the two boundaries that is represented by a membership function.

$$A = \int_x \mu_A(x)/x \qquad (1)$$

Where x is an element in X and $\mu_A(x)$ is a membership function. A Fuzzy set is characterized by a membership function that has grades between the interval [0, 1] called grade membership function. There are different types of membership function, namely, triangular, trapezoidal, Gaussian etc.

Fuzzy logic consists of the following three stages:
    1. Fuzzification
    2. Inference Engine
    3. Defuzzification

The Fuzzifier transforms the inputs into a membership value for the linguistic terms. The function of inference engine is to develop the complexity matrix for producing a new linguistic term to determine the productivity rate by using fuzzy rules. A defuzzifier carries out the Defuzzification process to combine the output into a single label or numerical value as required.

### 3.2 Fuzzy Analogy

Fuzzification of classical analogy procedure is Fuzzy analogy. It comprises the following procedures, viz., 1) Identification of cases, 2) Retrieval of similar cases and 3) Case adaptation. Each step is the fuzzification of its equivalent classical analogy procedure.

#### 3.2.1 Identification of cases

The goal of this step is the characterization of all software projects by a set of attributes. Selecting attributes, which will describe software projects, is a complex task in the analogy procedure. Indeed, the selection of attributes depends on the objective of the CBR system. In this case, the objective is to estimate the software project effort. Consequently, the attributes must be relevant for the effort estimation task. The objective of the proposed Fuzzy Analogy approach is to deal with categorical data. So, in the identification step, each software project is described by a set of selected attributes which can be measured by numerical or categorical values. These values will be represented by fuzzy sets.

In the case of numerical value $x_0$, its fuzzification will be done by the membership function which takes the value of 1 when $x$ is equal to $x_0$ and 0 otherwise. For categorical values, $M$ attributes are considered and for each attribute $M_j$, a measure with linguistic values is defined ($A_k^j$). Each linguistic value $A_k^j$ is represented by a fuzzy set with a membership function ($\mu_{A_k^j}$).

It is preferable that these fuzzy sets satisfy the normal condition. The use of fuzzy sets to represent categorical data, such as 'very low' and 'low', is similar to how humans interpret these values and consequently it allows dealing with imprecision and uncertainty in the case identification step.

#### 3.2.2 Retrieval of Similar Cases

This step is based on the choice of software project similarity measure. In this method, a set of candidate measures for software project similarity has been proposed for software project similarity. These measures assess the overall similarity of two projects $P_1$ and $P_2$, $d(P_1, P_2)$ by combining all the individual similarities of $P_1$ and $P_2$ associated with the various linguistic variables $V_j$ describing the project $P_1$ and $P_2$, $d_{V_j}(P_1, P_2)$. After an axiomatic validation of some proposed candidate measures for the individual distances $d_{V_j}(P_1, P_2)$, two measures have been retained [14].

$$d_{V_j}(P_1, P_2) = \begin{cases} \max_k \min(\mu_{A_k^j}(P_1), \mu_{A_k^j}(P_2)) \\ \max-\min \ aggregation \\ \sum_k \mu_{A_k^j}(P_1) \times \mu_{A_k^j}(P_2) \\ sum-product \ aggregation \end{cases} \quad (2)$$

Where $A_k^j$ are the fuzzy sets associated with $V_j$ and $\mu_{A_k^j}$ are the membership functions representing fuzzy sets $A_k^j$. Scale factors (SF) are understanding product objectives, flexibility, team coherence, etc., Effort multipliers (EF) are software reliability, database size, reusability, complexity, etc. The imprecision of the cost drivers significantly affects the accuracy of the effort estimates which are derived from effort estimation models. Since the imprecision of software effort drivers cannot be overlooked, a fuzzy model gains advantage in verifying the cost drivers by adopting fuzzy sets.

$$Effort = A * (SIZE)^{B + 0.01 * \sum_{i=1}^{N} SF_i} * \prod_{i=1}^{N} EM_i \quad (3)$$

Where $A$ and $B$ are constants, SF is the scale factor and EM is effort multipliers. By using the above formula the effort is estimated. The cost drivers are fuzzified using triangular and trapezoidal fuzzy sets for each linguistic value such as very low, low, nominal, high etc. as applicable to each cost driver.

Rules are developed with cost driver in the antecedent part and corresponding effort multiplier in the consequent part. The defuzzified value for each of the effort multiplier is obtained from individual Fuzzy Inference Systems after matching, inference aggregation and subsequent Defuzzification. Total Effort is obtained after multiplying them together. The high values for the cost drivers lead an effort estimate that is more than three times the initial estimate, whereas low values reduce the estimate to about one third of the original. This highlights the vast differences between different types of projects and the difficulties of transferring experience from one application domain to another

3.2.3 Case adaptation

The objective of this step is to derive an estimate for the new project by using the know effort values of similar projects. We are not convinced in fixing the number of analogies in this step. In our proposed method, all the projects in a dataset are used to derive the new project estimate. Each historical project will contribute, in the calculation of the effort of the new project according to the similarity.

## 4. Results and Discussion

The datasets used in the study is the Desharnais dataset [15], NASA 93 [12] and COCOMO NASA dataset shown in Table 1 and Table 2.

Table 1: Desharnais project features

| Variable | Description | Type |
|---|---|---|
| ExpEquip | Experience of Equipment | Continuous |
| ExpProj Man | Experience of Project Manager | Continuous |
| Trans | Transactions | Continuous |
| Raw FP | Raw Function Points | Continuous |
| Adj. Factor | Technology Adjustment Factor | Continuous |
| Adj.FP | Adjusted Function Points | Continuous |
| Dev Env | Development Environment | Categorical |
| Year Fin | Year Finished | Continuous |
| Entities | Number of entities | Continuous |
| Effort | Actual effort | Continuous |

Table 2: Nasa93 project features

| Variable | Description | Type |
|---|---|---|
| Acap | Analysts Capability | Categorical |
| Pcap | Programmers Capability | Categorical |
| Aexp | Application Experience | Categorical |
| Modp | Modern Programming Practices | Categorical |
| Tool | Use of Software tools | Categorical |
| Vexp | Virtual Machine Experience | Categorical |
| Lexp | Language Experience | Categorical |
| Sced | Schedule Constraint | Categorical |
| Stor | Main Memory Constraint | Categorical |
| Data | Database Size | Categorical |
| Time | Time Constraint for CPU | Categorical |
| Turn | Turnaround time | Categorical |
| Virt | Machine Volatility | Categorical |
| Rely | Required Reliability | Categorical |
| Cplx | Process Complexity | Categorical |
| Loc | Line of Code | Continuous |
| DevEff | Development Effort | Continuous |

Table 3, summarizes the number of projects collected under each dataset with the max effort which is compared with the estimated method. From this table, it is inferred that the datasets have a very low max effort when compared to the actual max effort for the proposed method.

Table 3 Comparison of actual max effort
With estimated max effort

| *Datasets* | *No of projects* | *Actual Max Effort* | *Estimated Max Effort* |
|---|---|---|---|
| Nasa60 | 60 | 3240 | 1594 |
| Desharnais | 77 | 23,940 | 10,950 |
| Nasa93 | 93 | 8211 | 1290 |

Figure.1 represents the graphical plot of the datasets versus the max effort of the estimated and the proposed method.

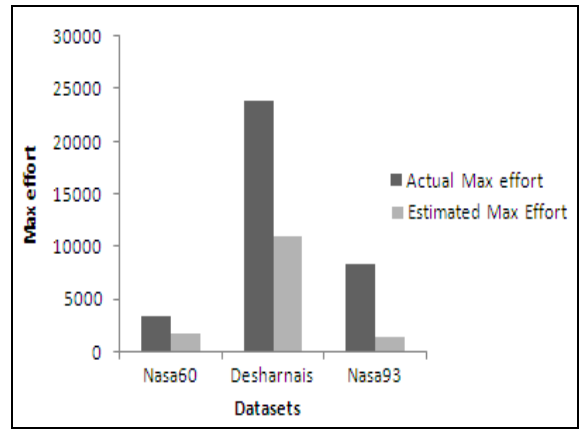

Figure 1: Comparative results of actual and estimated maximun efforts

In this method, 30% of test data sets is taken from NASA60, NASA93 and Desharnais to measure the average effort in comparison with the actual values. From the comparative results given in Table 4, it is predicted that the existing average effort using the selected features is very high compared to the proposed method for each dataset while considering the full-fledged features sets.

Table 4. Comparative results of average effort and actual effort

| Datasets | Test set | Actual Avg Effort | Existing Method | | Proposed Method | |
|---|---|---|---|---|---|---|
| | | | *No. of Feature* | *Avg. Effort* | *No. of Feature* | *Avg. Effort* |
| Nasa60 | 16 | 340.49 | 3 | 354.66 | 16 | 284.34 |
| Desharnais | 22 | 5119.3 | 2 | 4852.17 | 10 | 2428.9 |
| Nasa93 | 26 | 734.03 | 4 | 722.96 | 16 | 640.65 |

Figure 2 represents the graphical plot of the datasets versus the average effort among the existing and the proposed method.

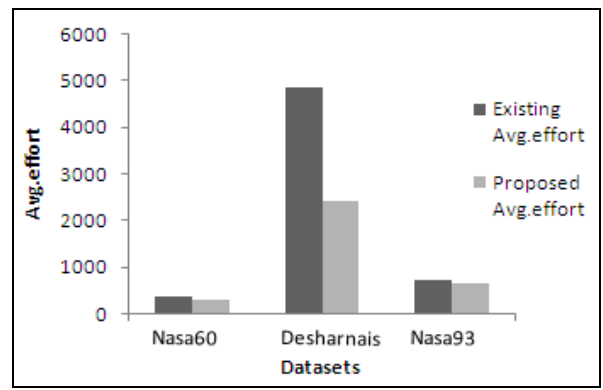

Figure 2: Comparative results of avg. effort in existing and proposed method

Therefore, it is concluded that by taking all the features into consideration for all the datasets, the average effort will be low in the proposed method in comparision to the existing method [4].

## 5. Conclusions

A new classical approach has been proposed in this paper to estimate the average software project effort. This approach is based on reasoning by analogy, fuzzy logic and linguistic quantifiers, which can be effectively used when the software projects are described by categorical and or numerical data. The new approach improves the classical analogy procedure while using the categorical data. In the fuzzy analogy approach, both categorical and numerical data are represented by fuzzy sets. The advantage of this method is that it can handle the imprecision and the uncertainty quite vividly while describing the software project. From the implementation of the results, it is observed that the proposed method has effectively estimated the average effort for the software project datasets.